# Population Metrology of a Hidden Exciton Reservoir: Quasi-Thermalization versus Localization

Guangyu Dai,[1] Xinyu Zhang,[1] Jiaxin Yu,[1,2,*] Zi'ang Xing,[1] Miao Cai,[1,2] Zhaoqi Gu,[3] Junyuan Zhang,[1] Lin Dou,[3] Fuxing Gu,[3] Yiming Zhu,[1,2,4] and Songlin Zhuang[1,2]

[1]*Terahertz Technology Innovation Research Institute, and Shanghai Key Lab of Modern Optical System, University of Shanghai for Science and Technology, Shanghai 200093, China*

[2]*National Key Laboratory of Terahertz Perception and Communication (TPCL)*

[3]*Laboratory of Integrated Opto-Mechanics and Electronics, University of Shanghai for Science and Technology, Shanghai 200093, China*

[4]*Shanghai Institute of Intelligent Science and Technology, Tongji University, Shanghai 200092, China*

Long-lived dark states can dominate the lowest-energy manifold of optically driven quantum materials, yet their occupation remains difficult to quantify, leaving it unclear whether it reflects thermal redistribution or kinetic trapping. We combine microsphere-enabled far-field access with quantitative optical-response calibration to retrieve dark-to-bright population ratios in monolayer $WSe_2$. Temperature-dependent measurements and controlled defect enhancement separate mobile and localized contributions. Near room temperature, the mobile dark-to-bright ratio reaches approximately 65% of its Boltzmann limit, indicating substantial but incomplete quasi-thermalization, whereas the low-temperature excess is dominated by defect-assisted localization. Dark-state dominance alone therefore does not establish equilibration, a distinction essential for interpreting transport and collective phases in optically hidden quasiparticle reservoirs.

Optically dark states often form the lowest-energy, long-lived sector of an excited quantum material and can store a substantial fraction of its photoexcited population [1–3]. Whether this population belongs to a mobile, thermalizing ensemble or is kinetically trapped in localized states is central to its physical interpretation: only the former can contribute to long-range exciton transport and interaction-driven phases [4,5], whereas the latter can mimic occupation of the lowest-energy manifold without establishing equilibration [6]. This distinction is especially important in assessing equilibrium or quasi-equilibrium excitonic condensation, where low-energy population buildup must be interpreted together with reservoir identity, thermalization, and coherence [4,5].

Spin-forbidden dark excitons in W-based transition-metal dichalcogenides provide a representative example. They lie below the bright excitons by $\Delta E_{\mathrm{DB}} \approx 45$ meV and can dominate the low-energy exciton manifold [3,7]. Yet quantitative access requires both making their weak out-of-plane transition detectable and redirecting its predominantly large in-plane momentum ($k_{\parallel}$) emission into the far-field collection cone [Fig. 1(a)] [7]. Bright–dark mixing [3,8], optical antennas [9,10] and photonic out-coupling [11,12] can address these visibility barriers, but simultaneously modify the radiative rate ($\Gamma$) and/or collection efficiency ($\eta$). Consequently, since $I_{\mathrm{det}} \propto N\Gamma\eta$, the enhanced emission ($I_{\mathrm{det}}$) need not imply increased population ($N$), let alone reveal whether the observed occupation arises from thermal redistribution or kinetic trapping. Quantitative population metrology of the dark reservoir therefore remains an experimental challenge.

In this Letter, we address this challenge through response-calibrated far-field retrieval of the dark-to-bright population ratio in monolayer $WSe_2$. Microsphere-enabled $k$-space folding [Fig. 1(b)], combined with Green-tensor quasinormal-mode (QNM) calibration and spatial inversion, converts detected emission into this population metric under continuous-wave excitation. Temperature-dependent measurements before and after defect modification resolve two physically distinct contributions: phonon-assisted redistribution within the mobile bright–dark manifold and nonequilibrium accumulation in localized states. Their separation identifies the quasi-thermalizing mobile reservoir relevant to exciton transport and interaction-driven phases, while excluding defect-assisted storage from the thermodynamic interpretation of the total dark population.

The device consists of monolayer $WSe_2$ on single-crystal Au, with a $SiO_2$ microsphere placed on top as a local optical probe [Fig. 1(c)]. Under nonresonant continuous-wave excitation, the planar $WSe_2$/Au emission at 7 K is dominated by a broad localized-state background, with no distinct dark-exciton feature [Fig. 1(d)]. Although the Au substrate enhances the coupling of out-of-plane dark-exciton dipoles to surface plasmon polaritons [13], most of this emission remains at large $k_{\parallel}$, outside the collection cone of the objective. The microsphere folds these high-$k_{\parallel}$ components into propagating far-field modes [Fig. 1(b)]. We use ~6.5-μm-diameter microspheres whose free spectral range of 42−45 meV allows adjacent resonances to sample the bright- and dark-exciton spectral regions. At 7 K, two of the resonances are set near 1.737 and 1.692 eV, sampling the reported bright- and dark-exciton spectral regions, respectively [3,7]. Notably, only the lower-energy resonance becomes strongly pronounced [Fig. 1(d)].

*Contact author: yujiaxin@usst.edu.cn

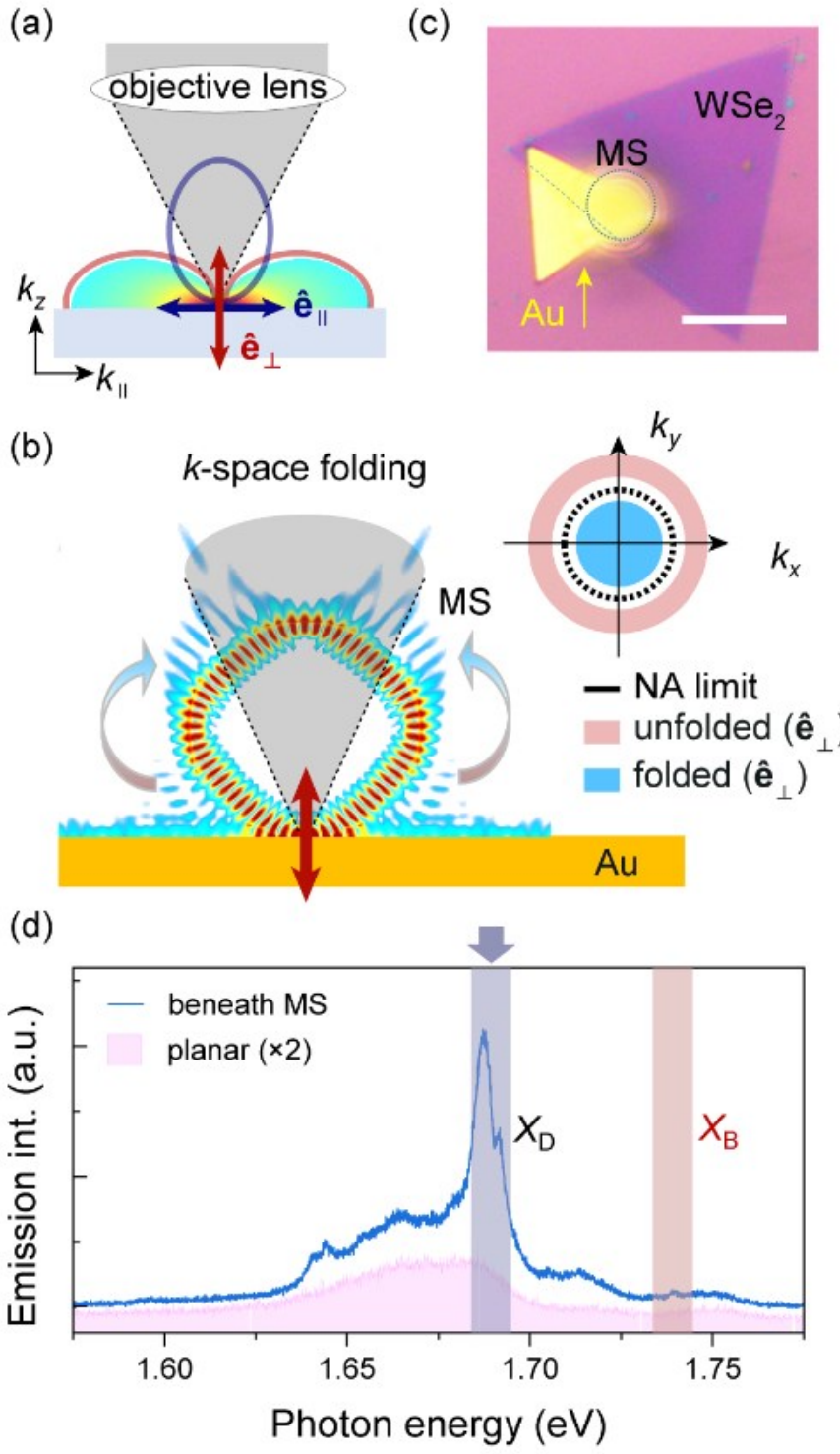

FIG. 1. Microsphere-enabled optical access to the dark-exciton reservoir. (a) Far-field radiation patterns of in-plane bright-exciton dipoles ($\hat{\mathbf{e}}_{\parallel}$, blue) and out-of-plane spin-forbidden dark-exciton dipoles ($\hat{\mathbf{e}}_{\perp}$, red). (b) Schematic of folding large-$k_{\parallel}$ out-of-plane emission into propagating far-field modes. Inset: unfolded (pink) and folded (blue) momentum distributions relative to the NA limit. (c) Optical micrograph of a representative device comprising a $SiO_2$ microsphere (MS) on monolayer $WSe_2$/Au. Scale bar, 10 μm. (d) Emission spectra at 7 K excited beneath the MS (blue) and on the adjacent planar region (pink; multiplied by 2). Shading marks the dark- ($X_D$) and bright ($X_B$)-exciton windows.

To identify the exciton reservoir probed by the lower-energy resonance, we compare its temperature dependence with the bright-exciton-dominated planar PL and with two control structures [16]. The $WSe_2$/Au resonance is only weakly correlated with planar PL ($r$ = 0.22) [Figs. 2(a) and 2(d)]. By contrast, it follows planar PL more closely in $MoSe_2$/Au ($r$ = 0.77) where the dark state lies slightly above the bright state ($\Delta E_{DB}$ ≈ −1.4 meV) [Figs. 2(b) and 2(e)] [14], and in $WSe_2/SiO_2$ ($r$ = 0.68), where coupling of out-of-plane dipoles to SPP modes is absent [Figs. 2(c) and 2(f)]. These controls respectively test the effects of bright–dark level ordering and dipole-selective SPP coupling. Together, they show that the dominant $WSe_2$/Au resonance is not a simple enhancement of planar bright-state emission, but is primarily sensitive to the out-of-plane dark-state reservoir.

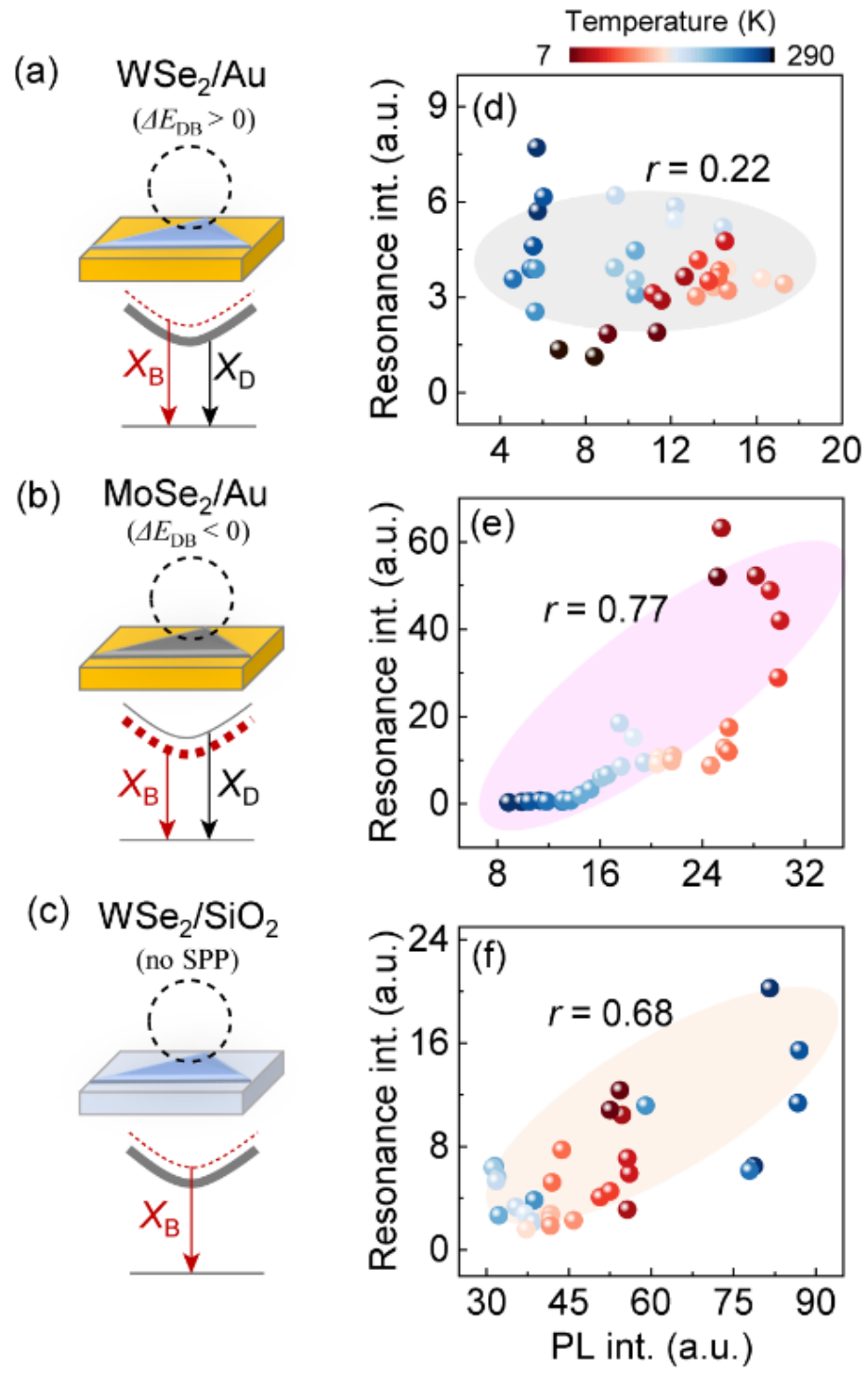

FIG. 2. Identification of the excitonic reservoir sampled by the microsphere resonance. (a–c) Exciton level ordering and optically accessible radiative channels in $WSe_2$/Au (a), $MoSe_2$/Au (b), and $WSe_2/SiO_2$ (c). Arrows denote microsphere-mediated far-field transitions from the bright ($|X_B\rangle$) and, where applicable, dark ($|X_D\rangle$) states to $|0\rangle$. (d–f) Integrated microsphere-resonance intensity versus bright-exciton-dominated planar PL for the corresponding structures in (a–c). Color encodes temperature; $r$ is the Pearson correlation coefficient.

Establishing the dominant state character of the resonance does not by itself determine the underlying population, because the detected intensity depends on both the radiative rate and the collection efficiency. We therefore express the measured signal as

$$I_{\mathrm{det}} = a(N_{\parallel}R_{\parallel} + N_{\perp}R_{\perp}), \quad (1)$$

where $a$ accounts for the common instrumental response and $N_{\perp}$ and $N_{\parallel}$ denote the populations sampled by the out-of-plane and in-plane dipole channels, respectively. For a unit dipole of orientation $i = \parallel, \perp$, located at $\mathbf{r}$ and emitting at angular frequency $\omega$, the detected response is $R_i(\mathbf{r}, \omega) = \eta_i(\mathbf{r}, \omega)\, \Gamma_i(\mathbf{r}, \omega)$. In $WSe_2$, the in-plane and out-of-plane channels are dominated by bright and spin-forbidden dark excitons, respectively [7].

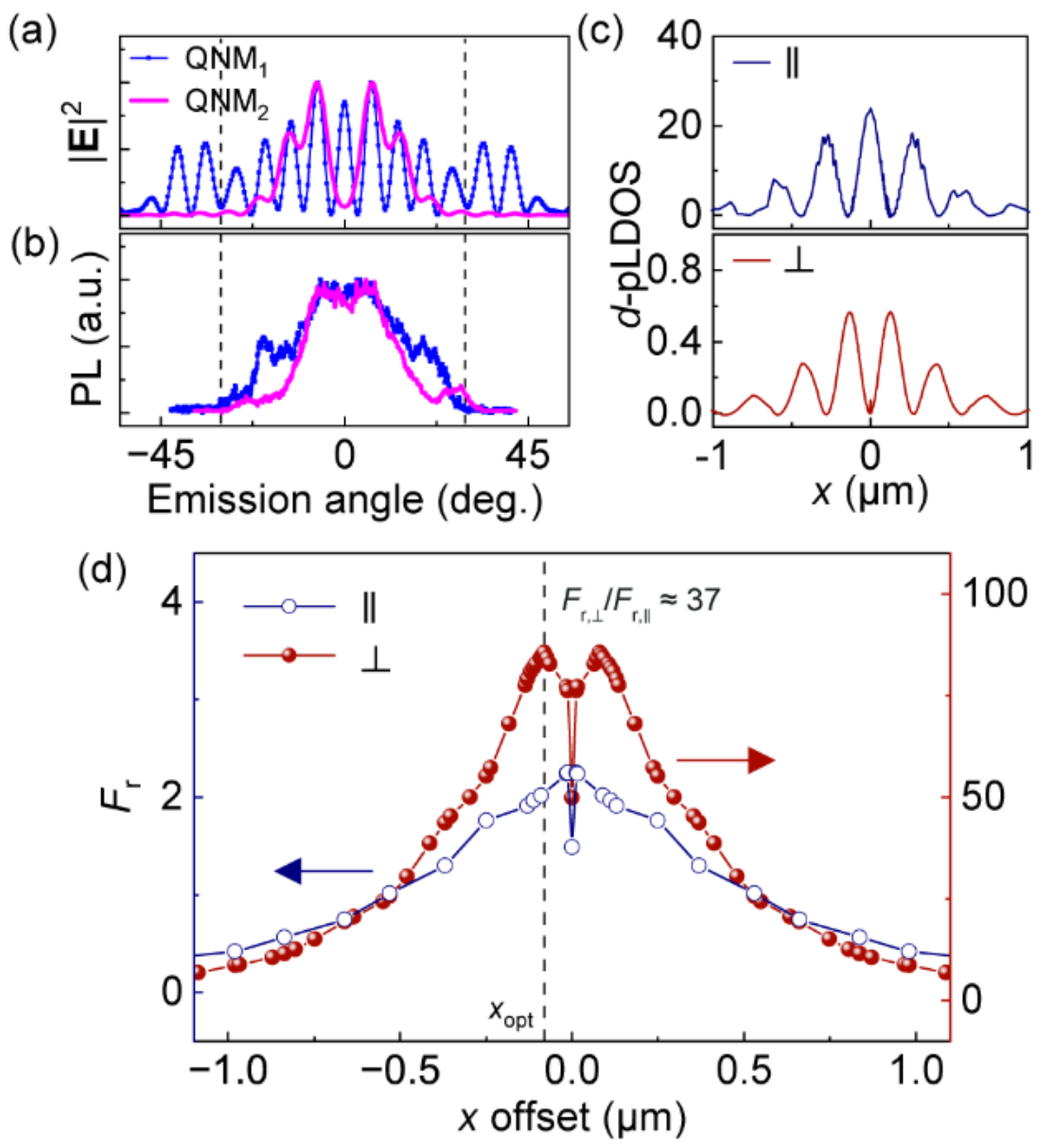

FIG. 3. Calibration of the orientation-dependent optical response. (a,b) Simulated emission intensities (a) and measured PL profiles at the corresponding spectral resonances (b) for the dominant $QNM_1$ and $QNM_2$. (c) $QNM_1$ contribution to the d-pLDOS along the $WSe_2$ plane for in-plane (∥, top) and out-of-plane dipoles (⊥, bottom). (d) Calculated radiative-rate modification factor $F_r$ versus lateral displacement ($x$) for in-plane (blue open symbols, left axis) and out-of-plane (red solid symbols, right axis) dipoles; solid curves guide the eye.

We determine $R_i$ using a Green-tensor QNM description of the microsphere–$WSe_2$/Au structure [15,16]. The model reproduces both the angular profiles of the two dominant resonances [Figs. 3(a) and 3(b)] and their displacement-dependent back-focal-plane evolution [16], thereby constraining the QNM contributions used to calculate the collection efficiencies. The same framework yields the dipole-projected local density of optical states (d-pLDOS) [Fig. 3(c)] and orientation-dependent radiative-rate modification ($F_r$). At the optimal position, preferential coupling of the out-of-plane dipole to the dominant QNM gives $F_{r,\perp}/F_{r,\parallel} \simeq 37$ [Fig. 3(d)], reducing the intrinsic bright–dark radiative-rate contrast and making the weak dark-channel emission accessible.

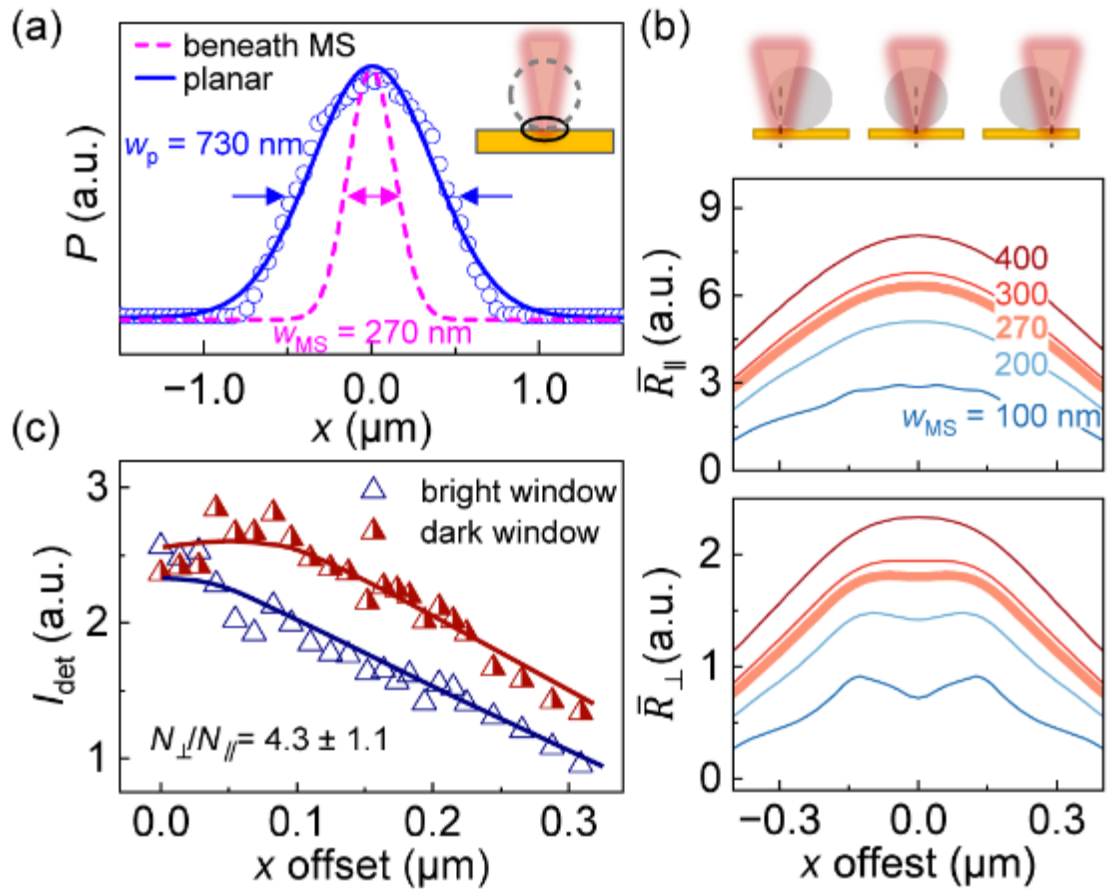

FIG. 4. Population retrieval from position-dependent optical responses. (a) Measured and fitted excitation profiles on planar $WSe_2$ (blue circles) and beneath the microsphere (magenta triangles), with beam-waist diameter $w_p = 730$ nm and $w_{MS} = 270$ nm, respectively. (b) Convolved position-dependent responses $\bar{R}_i(x)$ for different excitation profiles; insets show the excitation displacement relative to the microsphere. (c) Integrated emission versus position in the bright (1.669 eV, blue open triangles) and dark (1.625 eV, red half-filled triangles) windows; solid curves are simultaneous fits to Eq. (3).

For population reconstruction, we convolve the point-dipole responses with the measured excitation profile $P(x)$:

$$\bar{R}_i(x) = \int R_i(u)P(u-x)\,\mathrm{d}u, \quad (2)$$

where $u$ and $x$ are the lateral dipole position and displacement of the excitation spot relative to the microsphere center, respectively (the spectral-window label $\omega$ is suppressed for clarity). The measured and modeled excitation profiles are shown in Fig. 4(a). Importantly, $\bar{R}_{\parallel}(x)$ and $\bar{R}_{\perp}(x)$ exhibit nonproportional spatial dependences [Fig. 4(b)]. The position scan therefore supplies an independent constraint beyond the absolute emission intensity, making the two dipole contributions identifiable. The emission integrated over the dark- and bright-exciton spectral windows evolves differently with position. At room temperature, the dark window at 1.625 eV exhibits a broad plateau followed by a decay, whereas the bright window at 1.669 eV decreases approximately monotonically [Fig.

4(c)]. We simultaneously fit the two traces according to

$$I_{\rm det}^{\rm B(D)}(x) = A^{\rm B(D)}\,\bar{R}_{\parallel}(x) + B^{\rm B(D)}\,\bar{R}_{\perp}(x), \quad (3)$$

Here the superscript denotes the spectral window, while $A$ and $B$ are the fitted amplitudes of the in-plane and out-of-plane response functions, respectively. The fits yield $A^{\rm D} \simeq 0$ and $B^{\rm B} \simeq 0$ within uncertainty [16], confirming weak cross-channel leakage and the spectral selectivity assumed in the population inversion.

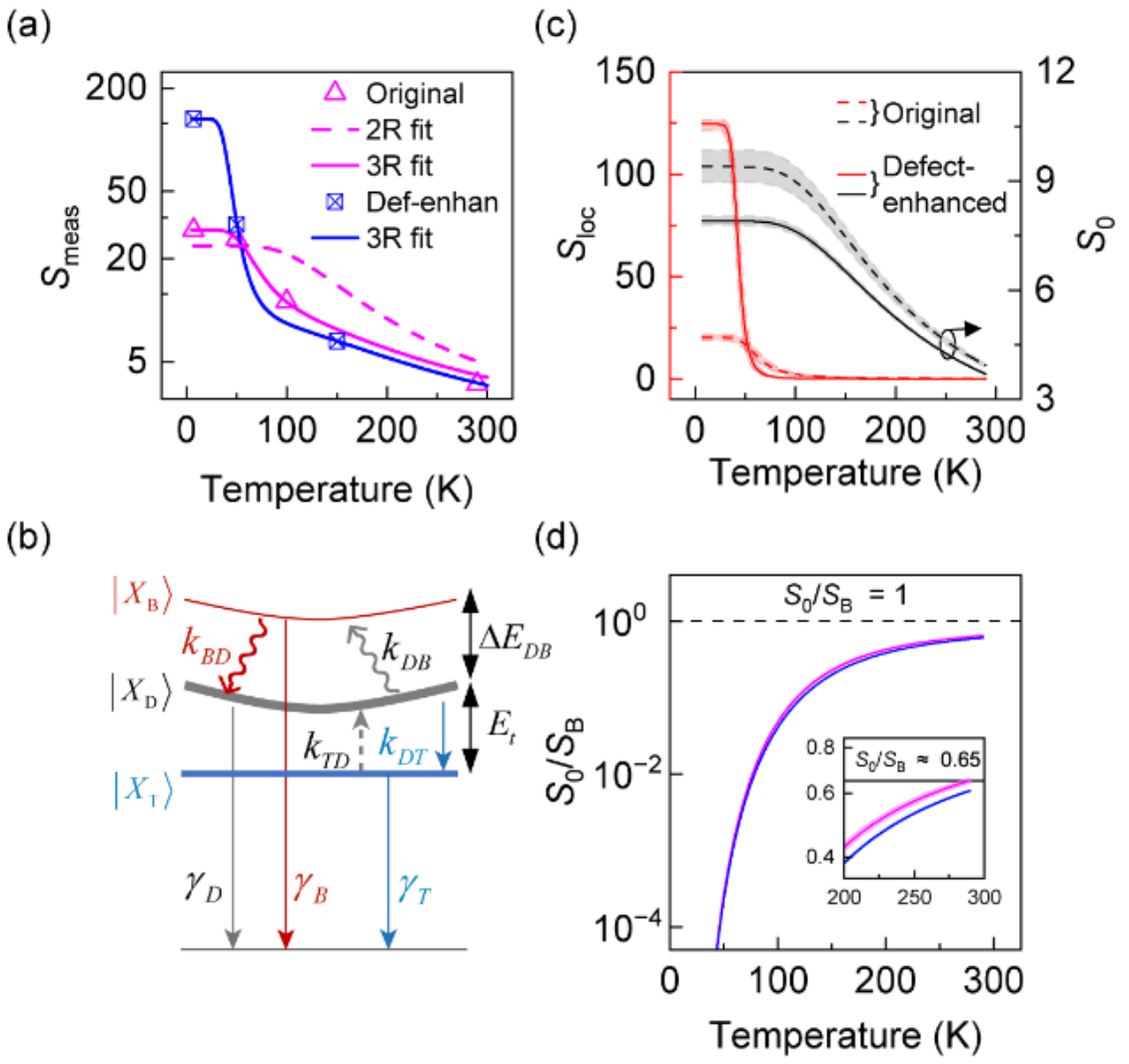

FIG. 5. Separation of mobile and localized dark-exciton populations. (a) Measured out-of-plane-to-bright population ratio ($S_{\rm meas}$) versus temperature before (magenta triangles) and after (blue squares) defect enhancement. Dashed and solid curves are independent two-reservoir (2R) and three-reservoir (3R), respectively. (b) Three-reservoir kinetic model with bright-exciton $N_{\rm B}$, mobile dark-exciton $N_{\rm D}^{\rm mob}$, and localized population $N_{\rm T}$. Arrows denote phonon-assisted exchange ($k_{\rm BD}$ and $k_{\rm DB}$), capture into localized states ($k_{\rm DT}$), thermally activated release ($k_{\rm TD}$), and irreversible loss ($\gamma_{\rm B}$, $\gamma_{\rm D}$, $\gamma_{\rm T}$). (c) Extracted mobile ($S_0$) and localized ($S_{\rm loc}$) contributions for the two datasets; dashed and solid curves denote the original and defect-enhanced datasets, respectively. (d) Mobile population relative to the Boltzmann limit ($S_0/S_{\rm B}$) versus temperature for both datasets; inset enlarges the near-room-temperature range.

The fitted amplitudes give the response-calibrated ratio $N_{\perp}/N_{\parallel} = B^{\rm D}/A^{\rm B} = 4.3 \pm 1.1$. Because each QNM samples only a finite fraction of the broader exciton emission band, we further correct for the calculated spectral sampling efficiencies [16]. This gives a room-temperature dark-to-bright population ratio

$$S_{meas} \equiv N_{\rm D}^{\rm tot} / N_{\rm B} \simeq 3.70 \pm 0.95, \quad (4)$$

Here $N_{\rm D}^{\rm tot}$ and $N_{\rm B}$ denote the spectrally corrected populations assigned to the out-of-plane dark-state and in-plane bright-state reservoirs, respectively. For the temperature series, resonance shifts and spectral sampling efficiencies were accounted for at each temperature, while the QNM angular and collection responses remained unchanged within experimental uncertainty [16].

We next determine whether this out-of-plane reservoir consists predominantly of mobile dark excitons participating in bright–dark exchange or contains an additional population accumulated in localized states. Exciton–phonon scattering mediates bright-to-dark relaxation in monolayer TMDs [17,18], while first-principles Boltzmann calculations show that the overall exciton distribution approaches thermal form more rapidly at elevated temperature [19]. We therefore compare the measured ratio $S_{\rm meas}$ with a phonon-assisted two-level exchange model previously applied to temperature-dependent $WSe_2$ populations [Fig. 5(a)] [20]. We describe bright-to-dark relaxation by $k_{\rm BD} = k_0(1 + n_{\rm ph})$ and dark-to-bright activation by $k_{\rm DB} = k_0 n_{\rm ph}$, where $n_{\rm ph}$ is the occupation of the effective phonon mode mediating the exchange. Assuming negligible direct generation of the dark reservoir under our excitation conditions, the steady-state mobile population ($N_{\rm D}^{\rm mob}$) ratio is

$$S_0(T) \equiv \frac{N_{\rm D}^{\rm mob}}{N_{\rm B}} = \frac{k_{\rm BD}(T)}{\gamma_D + k_{\rm DB}(T)} = \frac{1+n_{ph}}{\beta + n_{ph}}, \quad (5)$$

where $\beta \equiv \frac{\gamma_{\rm D}}{k_0}$, and $\gamma_{\rm D}$ is the decay rate of dark excitons. The two-level model describes the common high-temperature trend but fails when fitted over the full temperature range [16]. In particular, below approximately 50 K, the measured out-of-plane population saturates far above the value predicted for the mobile bright–dark manifold [16]. As $n_{\rm ph} \rightarrow 0$, the thermally activated dark-to-bright channel is suppressed; this excess cannot be assigned solely to unrestricted bright–dark equilibration. This discrepancy requires an additional low-temperature reservoir beyond the mobile bright–dark manifold. We therefore introduce a localized reservoir ($N_{\rm T}$), coupled to the mobile dark population through capture and thermally activated release [Fig. 5(b)]; the corresponding steady-state rate equations are given in Ref. 16. Here, $k_{\rm DT}$ is the capture rate from the mobile dark reservoir into localized states, whereas $k_{\rm TD}$=$k_{\rm TD}^0 \exp(-E_{\rm t} / k_{\rm B}T)$ describes thermally activated release. The measured ratio is then $S_{meas} \equiv (N_{\rm D}^{\rm mob} + N_T) / N_{\rm B} = S_0 + S_{loc}$, where $S_0$ is the mobile contribution governed primarily by bright–dark exchange, and $S_{loc}$ is the localized contribution. The model reproduces both the common high-temperature decrease and the low-temperature saturation [Figs. 5(a) and 5(c)]. The extracted decomposition shows that the localized contribution becomes dominant at low temperature but is progressively depleted as thermal release becomes effective.

We test this assignment by intentionally enhancing defect-related localization by using strong irradiation and repeating the population measurement [16]. Independent fits of the three-reservoir model to the original and defect-enhanced datasets, with all parameters allowed to vary, yield $E_t = 32.0$ meV and $\beta = 0.107$ for the original sample, and $E_t = 40.2$ meV and $\beta = 0.126$ after defect enhancement. Despite the fully independent fits, the two extracted mobile contributions $S_0(T)$ differ by only approximately 6% near room temperature, while the activation energies remain on a comparable tens-of-meV scale. In contrast, the model-extracted localized-reservoir amplitude and thermally activated release prefactor increase by factors of approximately 6.2 and 146, respectively [16]. Together with the enhanced low-temperature branch and convergence of the two datasets at high temperature [Fig. 5(a)], this selective response indicates that defect enhancement primarily modifies the weight and kinetics of the localized reservoir, while leaving the mobile bright–dark exchange branch comparatively stable.

Having separated the localized contribution, we finally compare the mobile component with the thermal Boltzmann ratio. For the valley-summed bright and mobile-dark populations, the common $K_\pm$ degeneracy cancels ($g_D/g_B = 1$), giving the Boltzmann-limit population ratio [17]

$$S_B(T) = \exp(\Delta E_{DB}/k_B T), \quad (6)$$

For the exchange model, the departure from this limit can be written as

$$\frac{S_0}{S_B} = \frac{n_{ph}}{\beta + n_{ph}}, \quad (7)$$

When exchange is rapid compared with irreversible dark-state loss, $n_{ph} \gg \beta$, this ratio approaches unity and detailed balance is recovered within the mobile manifold [18]. At low temperature, $n_{ph} \ll \beta$, and loss competes effectively with the activated exchange required to establish equilibrium. The extracted $S_0/S_B$ increases continuously with temperature and reaches ~0.65 near room temperature [Fig. 5(d)], demonstrating a substantial but incomplete approach to the bright–dark detailed-balance limit within the mobile manifold. Accordingly, within the investigated temperature range, the near-room-temperature mobile reservoir provides the strongest population-level basis for a thermodynamic interpretation of collective behavior, whereas the large low-temperature dark population is dominated by localization. Although this does not by itself establish condensation, it identifies the mobile room-temperature reservoir as the relevant population for testing dark-exciton condensation scenarios [4] and for interpreting recent phase-coherence measurements in a related $WSe_2$/Au microsphere architecture [21].

In summary, we establish calibrated population-level metrology for the dark-exciton reservoir in monolayer $WSe_2$. Combining temperature-dependent population retrieval with controlled defect modification reveals two physically distinct origins of dark-state dominance: the mobile bright–dark manifold reaches approximately 65% of its Boltzmann-limit population ratio near room temperature, whereas the large low-temperature dark population is dominated by defect-assisted localization. Thus, a large occupation of the lowest-energy exciton channel alone is not evidence of equilibration; both population calibration and reservoir separation are required to determine whether it has a thermodynamic origin.

This work was supported by National Natural Science Foundation of China (62588201, 12074259, 62122054, 62422510 and U24A20226), the 111 project (D18014), Major Basic Research Program of Shandong Provincial Natural Science Foundation (ZR2026ZD26), National Key Research and Development Program of China (2023YFF0719200, 2023YFF0717900).

G. D. and X. Z. contributed equally to this work.